\documentclass[12pt,preprint]{aastex}
\shorttitle{Multiwavelength observations of a TeV-Flare from W Comae}
\shortauthors{a}

\def\lesssim{\lower4pt\hbox{${\buildrel < \over \sim}$}}
\def\gtrsim{\lower4pt\hbox{${\buildrel > \over \sim}$}}

\begin{document}

\title{Multiwavelength observations of a TeV-Flare from W Comae}

\author{
V.~A.~Acciari\altaffilmark{1},
E.~Aliu\altaffilmark{3},
T.~Aune\altaffilmark{4},
M.~Beilicke\altaffilmark{5},
W.~Benbow\altaffilmark{1},
M.~B{\"o}ttcher\altaffilmark{6},
D.~Boltuch\altaffilmark{3},
J.H.~Buckley\altaffilmark{5},
S.~M.~Bradbury\altaffilmark{7},
V.~Bugaev\altaffilmark{5}
% ...
}

\altaffiltext{1}{Fred Lawrence Whipple Observatory, Harvard-Smithsonian Center for Astrophysics, Amado, AZ 85645, USA}
\altaffiltext{3}{Department of Physics and Astronomy and the Bartol Research Institute, University of Delaware, Newark, DE 19716, USA}
\altaffiltext{4}{Santa Cruz Institute for Particle Physics and Department of Physics, University of California, Santa Cruz, CA 95064, USA}
\altaffiltext{5}{Department of Physics, Washington University, St. Louis, MO 63130, USA}
\altaffiltext{6}{Astrophysical Institute, Department of Physics and Astronomy, Ohio University, Athens, OH 45701}
\altaffiltext{7}{School of Physics and Astronomy, University of Leeds, Leeds, LS2 9JT, UK}
% ...

\begin{abstract}
We report results from an intensive multiwavelength campaign on the intermediate-frequency-peaked
BL Lacertae object W Com (z=0.102) during a strong outburst of very high energy gamma-ray emission
in June 2008. 
The very high energy gamma-ray signal was detected by VERITAS on 2008 June 7-8 with a flux 
F($>200$ GeV) $= (5.7\pm0.6)\times 10^{-11}$ cm$^{-2}$s$^{-1}$, about three
times brighter than during the discovery of gamma-ray emission from W Com by VERITAS in 2008 March.
The initial detection of this flare by VERITAS at energies above 200 GeV
was followed by observations in
high energy gamma-rays (AGILE, $E_{\gamma}\ge$ 100 MeV), and X-rays (Swift and XMM-Newton),
 and at UV, 
 and ground-based optical and radio monitoring through the GASP-WEBT consortium  and other observatories.
Here we describe the multiwavelength data and derive the spectral energy distribution (SED) 
of the source from contemporaneous data taken throughout the flare.
\end{abstract}

\keywords{BL Lacertae objects: individual (W Com) - gamma rays: observations}

\section{Introduction}

W Com (ON 231; $z=0.102$) is a gamma-ray blazar classified as an intermediate-frequency-peaked BL Lac (IBL) object
\citep{Nieppola-2006},
based on the locations of its low-energy synchrotron peak and high-energy peak in its spectral energy distribution (SED). 
The majority of the blazars detected at very high energies (VHE, $E>100$ GeV) by ground-based imaging atmospheric
 Cherenkov telescopes (IACTs) are high-frequency peaked BL Lacs (HBL), 
 characterized by synchrotron peaks in the X-ray band (often at energies of $\sim 100$ keV). 
 Due to the improved sensitivity of current-generation IACTs such as VERITAS, IBLs are 
 attractive targets of observations at VHE gamma rays, particularly because they offer the possibility of extension
 of the VHE blazar catalog to include non-HBLs.
 VHE observations of different blazar classes, including flat-spectrum radio quasars (FSRQs) 
 and BL Lac objects, will help in our understanding of the relationship of the different
  blazar populations and, ultimately, the mechanism for particle acceleration and emission in the 
  highly-relativistic jets. 

W Com was the first IBL to be detected at very high energies \citep{Acciari2008b}.
It was discovered as a VHE source by VERITAS during observations carried out over
a four month period in 2008 (Jan to Apr).
During this time a strong gamma-ray outburst was measured over a 4-day interval, 
when the source flared in the middle of March. 
VERITAS reported a steep photon spectrum $(\Gamma=3.81\pm0.35_{\rm stat}\pm0.34_{\rm sys})$\footnote{The subscripts
\emph{stat} and \emph{sys} denote the statistical and systematic error.} and
an integral flux of 9\% of the Crab Nebula flux during the flare nights. 
The VERITAS detection triggered \emph{Swift} observations, and the multiwavelength data 
obtained were  adequately explained with a synchrotron-external Compton (EC)
 leptonic model \citep{Acciari2008b}. 

In this article we report on a second VHE flare in W Com observed by VERITAS in 2008 June. During this flare, when the source was approximately three times brighter than during the 2008 March observations, a multiwavelength campaign was triggered, including observations with the space-based AGILE gamma-ray telescope and the \emph{Swift} and \emph{XMM-Newton} X-ray telescopes.
Here we describe the multiwavelength data and derive the spectral energy distribution (SED) of the source from contemporaneous data taken
throughout the flare.

\section{Observations and Results}

\begin{figure}
\includegraphics[height=0.6\paperheight]{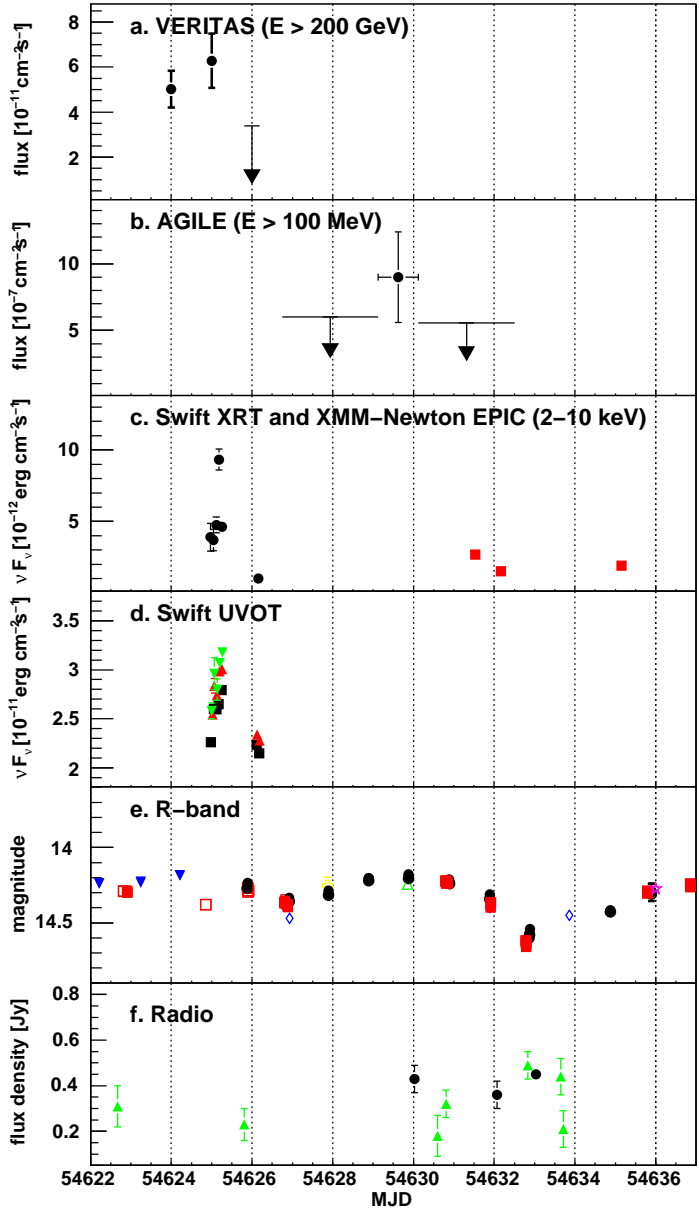}
\caption{\label{fig-lightcurve}
Multiwavelength light curve  of W Com for MJD 54622 to 54636.
Panel a: VHE gamma-ray light curve (E$>$200 GeV) as measured by VERITAS. 
The flux in VHE gamma rays corresponds to approximately 25\% of the flux of the Crab Nebula above 200 GeV.
Panel b: Gamma-ray light curve (E$>$ 100 MeV) as measured by AGILE.
Panel c: X-ray (\emph{Swift XRT}: 2-10 keV; circles; \emph{XMM-Newton EPIC}: squares) 
Panel d: Swift UVOT (UVW1: squares; UVM2: downward-pointing triangles; UVW2: upward-pointing triangles)
Panel e: Light curves of negative optical magnitudes
(R-Band; filled circles:  Tuorla; filled squares: Abastumani; filled triangles: San Pedro Martir; diamonds: Sapienza University;
 open circles: KVA; open squares: Crimean; open stars: NOT; open triangles: Torino; open crosses: Talmassons ).
Panel f: Radio light curve (circles: UMRAO 14.5 GHz; triangles: Mets\"ahovi 37 GHz).
Downward pointing
arrows indicate upper flux limits (99\% confidence level; \cite{Helene-1983}).
}
\end{figure}

A summary of the complete multiwavelength data set on W Com for observation times close to the VHE detection on 2008 June 7-9 
can be found in Table \ref{table:Observatories}
and Figure \ref{fig-lightcurve}.

\subsection{VERITAS: VHE Gamma-ray observations}

VERITAS is an array of four imaging Cherenkov telescopes
located at the Fred Lawrence Whipple Observatory
in southern Arizona.
It combines a large effective
area (up to $10^5$ m$^2$) over a wide energy range (100 GeV
to 30 TeV) with good energy resolution (15-20\%) and angular resolution
($\approx 0.1^{\mathrm{o}}$).
The field of view of the VERITAS telescopes is 3.5$^{\mathrm{o}}$.
The high sensitivity of VERITAS allows
the detection of sources with a flux of 1\% of the Crab
Nebula in less than 50 hours of observations.
For more details on the VERITAS
instrument, see \cite{Holder-2006} or \cite{Weekes-2002}.

VERITAS observed W Com for 230 minutes on 2008 June 7-9.
All observations pass quality-selection 
criteria, which remove data taken during bad weather or affected by
hardware-related problems.
The data were taken in wobble mode, wherein the source was positioned at a fixed
offset of 0.5$^{\mathrm{o}}$ in one of four directions (North, South, East, West)
from the camera center.
This
allows the simultaneous estimate of the background  \citep{Fomin-1994}.
The regions around the VHE gamma-ray blazar 1ES 1218+304 \citep{Acciari2008c},
located about 2$^{\mathrm{o}}$ north of W Com, and around bright stars (B-band magnitude brighter than 6)
are excluded from the background estimation.
All observations were undertaken in moonlight conditions, where 
the elevated background light levels lead to a lower sensitivity for the
detection of gamma rays at the threshold.
The threshold of the first-level trigger system \citep{Holder-2006}
was increased to 70 mV (compared to a
default value of 50 mV) to allow for very high background moonlight levels 
during observations on 2008 June 9. 
Table \ref{table:VERITASobservations} lists observation times, elevation range,
and background light conditions for the VERITAS observations.
The different elevations of observation combined with
the continuously changing background light conditions  
result in a range of energy thresholds from 200 to 420
GeV\footnote{
The energy threshold is defined as the energy at which the peak of the differential
counting rate for a Crab Nebula-like spectrum occurs.
}.

The analysis steps consist of calibration and integration of the flash-ADC traces, image cleaning,
second-moment parameterization of the telescope images \citep{Hillas-1985},
stereoscopic reconstruction of the event impact position and direction, 
gamma-hadron separation (see e.g., \cite{Krawczynski-2006}),
and 
the generation of photon maps.
Most of the far more numerous background events are rejected by comparing the
parameterized shape of the event images in each 
telescope with the expected shapes of gamma-ray showers modeled by Monte Carlo simulations.
\emph{Mean-reduced-scaled width} and \emph{mean-reduced-scaled length}
cuts (see definition in \cite{Acciari2008a}), and an additional cut on the arrival direction of the incoming gamma ray
 ($\Theta^2$, defined as the square of the angular distance to the position of W Com to the reconstructed shower direction),
reject more than 99.9\% of successfully reconstructed cosmic-ray background events
while keeping 45\% of the gamma rays.
The cuts applied here are:
integrated charge per image $>$75 photoelectrons,
 mean-reduced-scaled width and length between -1.2 and 0.5, and $\Theta^2 < 0.015$ deg$^2$.
The number of background events in the source region are estimated from the same field of view
using the ``reflected-region'' model with 10 background regions \citep{Aharonian-2001}.

The energy of each event is estimated from detailed Monte Carlo simulations of extensive air showers and the
response of the telescopes, focal plane detectors and electronics. 
The energy reconstruction algorithm uses lookup tables and 
determines the energy of an event as a function of impact parameter,
integrated charge per image, background light level, offset of the arrival direction from the center
of the camera, and zenith and azimuth angle. 
Gamma-ray collection areas for these different observing conditions are calculated using the same Monte Carlo
simulations \citep{Mohanty-1998}.
The finite energy resolution is taken into consideration by calculating collection areas as a function of
reconstructed energy.
The dependence of the collection area on the spectral index is taken into account by an iterative process, 
where collection areas are calculated using the spectral index obtained in the previous step.
Convergence is usually achieved after 2-3 steps.
The spectral reconstruction algorithm assigns to each event a collection area according to its estimated energy,
background light level, offset of the arrival direction from the center
of the camera, assumed spectral index, and zenith and azimuth angles. 
Varying conditions, like changing elevations or background light levels are therefore taken into account 
in the flux calculations and spectral energy reconstruction.
It should be noted that the definition of energy threshold used here implies that collection areas are non-zero 
below the stated threshold value. Gamma rays are collected, although with lower efficiency, at energies well below
200 GeV even for the brightest background light levels.
The systematic error  in the estimation of the gamma-ray energy is dominated by variabilities and uncertainties 
in the atmospheric conditions, overall Cherenkov photon collection efficiency, and limitations of the 
Monte Carlo simulations. 

\begin{figure}
\plotone{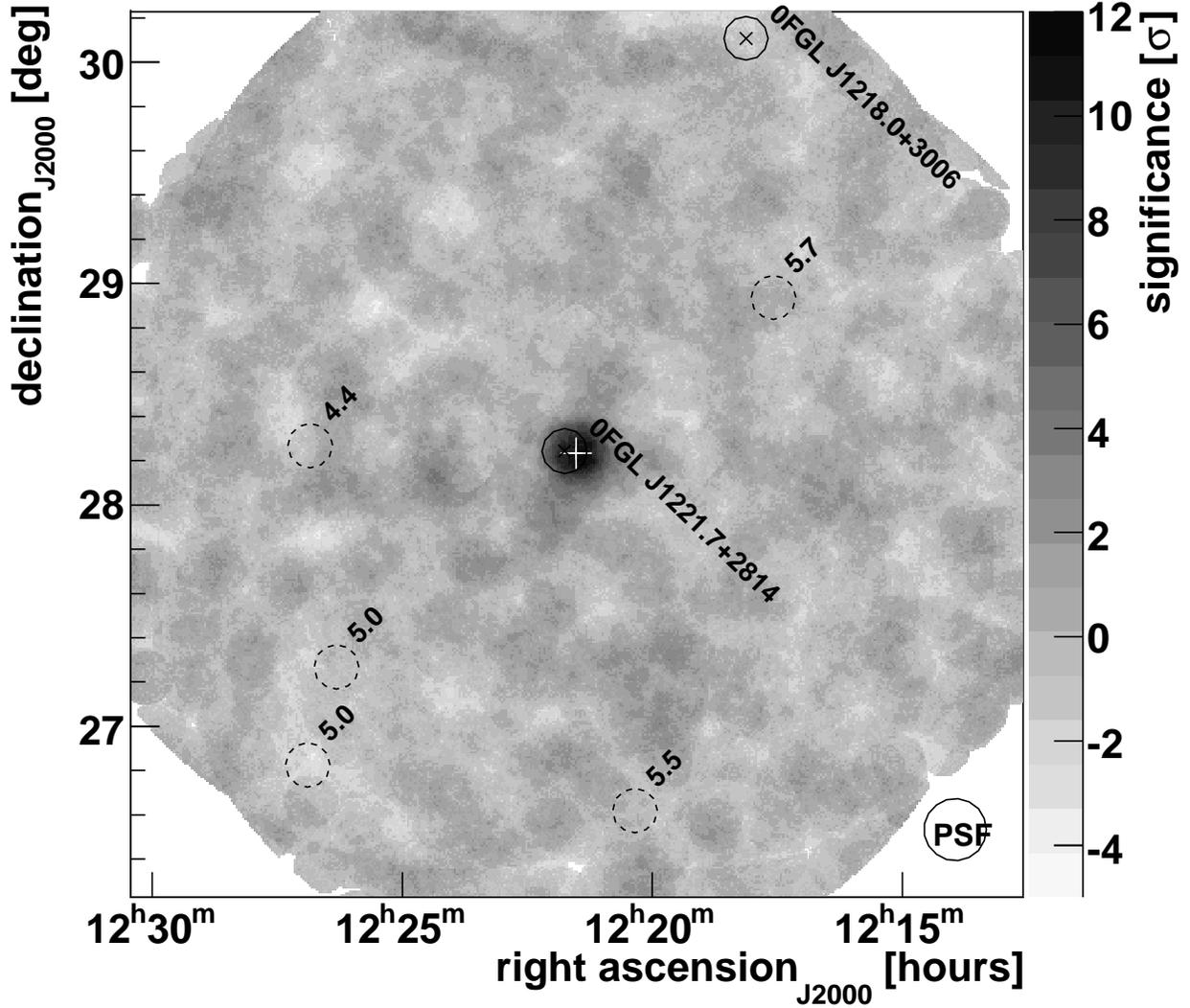}
\caption{\label{fig-skyplot}
Sky map of significances of gamma-ray emission from the region around W Com.
The background is estimated using the reflected region model  (10 background regions, oversampling
radius 0.12$^{\mathrm{o}}$). 
The position of W Com derived from radio data \citep{Fey-2004}
is indicated by a white cross.
The dashed circles indicate positions of bright stars and their B-band magnitudes in the field of view; regions around
these stars are excluded from the background estimation.
Two sources listed in the Fermi bright gamma-ray source list \citep{Abdo-2009}, and firmly associated with
the blazars W Com and B2 1215 \citep{Abdo-2009a}, are shown with their
95\% confidence area as circles with 'x' in their centre.
The circle at the bottom right indicates
the angular resolution of the VERITAS observations.
}
\end{figure}

Figure \ref{fig-skyplot} shows the sky around
W Com as seen by VERITAS in VHE gamma rays. 
A  significant flux of very-high-energy
gamma rays from W Com is detected by VERITAS
for the entire data set taken on 2008 June 7-9. 
A total of 117 excess events (195 \emph{on-source} events
and 78 normalized \emph{off-source} events, normalization factor of 0.10) 
are measured. 
This corresponds to a
significance of 10.3 standard deviations, 
calculated following Equation 17 in \cite{Li-1983}.
Table \ref{table:VERITASobservations} lists the
daily significances
and fluxes above 200 GeV, assuming a power-law like spectral shape
with a photon index of 3.68 (see next paragraph); Figure \ref{fig-lightcurve} shows
the light curve for these observations. 
W Com is not detected on 2008 June 9 (MJD 54626),
but observations were restricted to only 32 min due to very high
background light levels caused by the Moon.
The average flux on 2008 June 7-8 is 2.5-3 times higher
than during the gamma-ray flare from W Com in March 2008 
\citep{Acciari2008b}.
The position of the peak of the gamma-ray excess,
reconstructed by fitting a 2D-Gaussian function to the uncorrelated 
excess sky map, is in agreement with the position of the radio
source associated with W Com \citep{Fey-2004}: $\Delta_{RA}=40''\pm31''_{stat}$, $\Delta_{dec}=-55''\pm41.4''_{stat}$.
The systematic uncertainty on the pointing, verified with optical pointing monitors, is less
than $50''$.
The morphology of the excess is compatible with the distribution expected from a point source.

\begin{figure}
\plotone{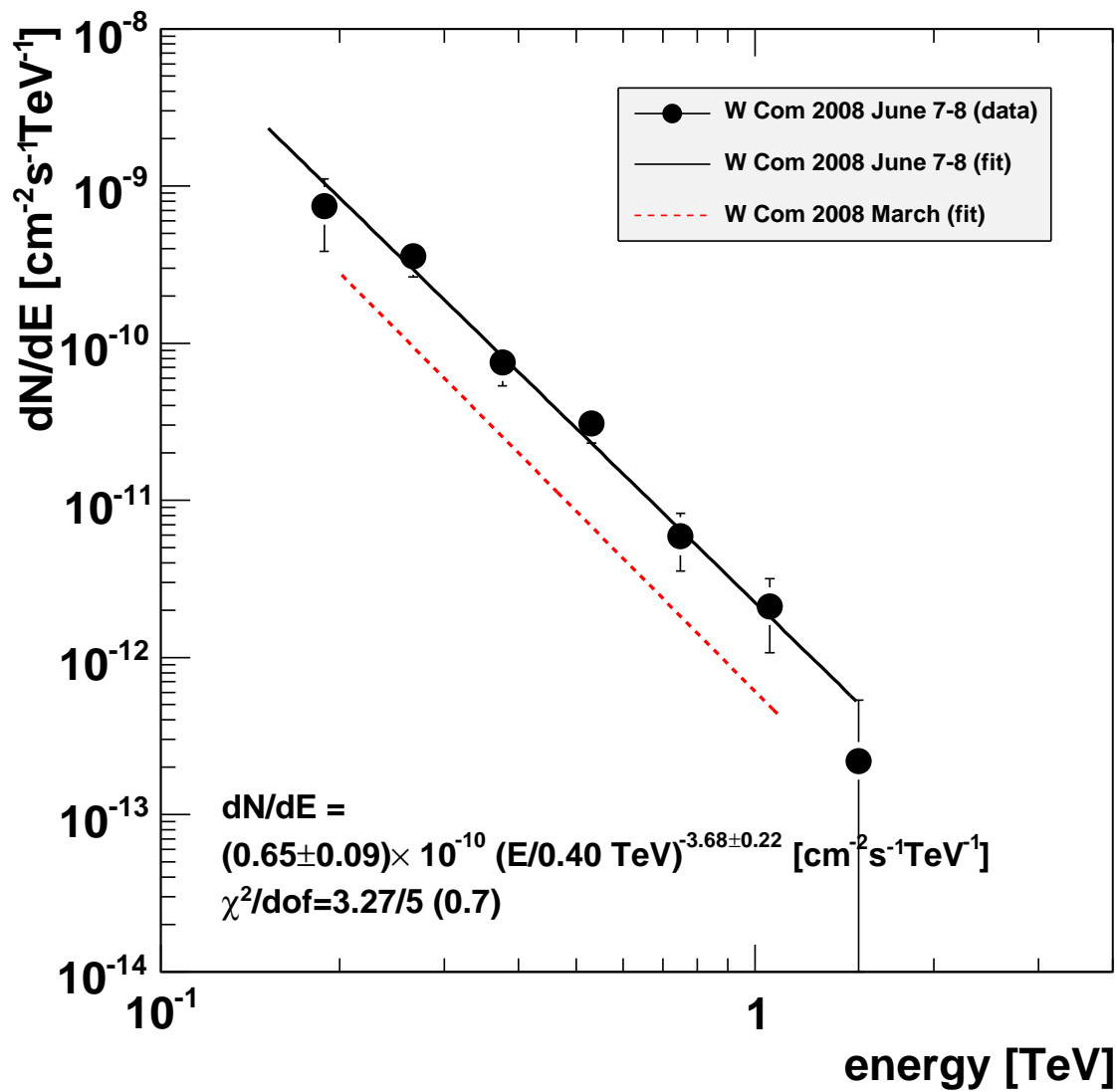}
\caption{\label{fig-espec}
Differential VHE photon spectrum  for W Com for MJD 54624.16
to 54625.24 (2008 June 7-8). 
The markers indicate measured data points and the continuous line a fit assuming a power-law
distribution. 
Error bars show statistical errors only.
For comparison, the photon spectrum of W Com derived from VERITAS measurements
in March 2008 \citep{Acciari2008b} is indicated by a dashed line.
}
\end{figure}

The differential photon spectrum between 180 GeV and
3 TeV for the measurements from 2008 June 7-8 is shown in Figure \ref{fig-espec}.
The shape of the spectrum is consistent with a power law
$\mathrm{dN/dE = C\times(E/400\ GeV)}$$^{-\Gamma}$ with an photon index
$\Gamma = 3.68\pm0.22_{stat}\pm0.3_{sys}$ and a flux normalization constant 
C = $(6.5\pm0.9_{stat}\pm1.3_{sys})\times10^{-11}$
cm$^{-2}$s$^{-1}$TeV$^{-1}$.
For comparison, the flare in VHE gamma rays from W Com in 2008 March \citep{Acciari2008b}
 is well fit
by a power law with a consistent $\Gamma = 3.81\pm0.35_{stat}\pm0.34_{sys}$, but smaller flux constant
C = $(2.00\pm0.31_{stat}\pm0.5_{sys})\times10^{-11}$
cm$^{-2}$s$^{-1}$TeV$^{-1}$.

\subsection{AGILE: HE Gamma-ray observations}

The Gamma-Ray Imaging Detector (GRID, 30 MeV - 30 GeV) onboard the high
energy astrophysics satellite AGILE \citep{Tavani2008a} pointed towards
W Com continuously from 2008 June 9 (18:00 UT) to 15 (12:00 UT) \citep{Verrecchia2008}.
The GRID data is analyzed using the AGILE standard pipeline \citep{Vercellone2008}, with a bin
size of $0.25^{\mathrm{o}} \times 0.25^{\mathrm{o}}$. Only events flagged as gamma rays and not recorded while
the satellite crossed the South Atlantic Anomaly are accepted.
Events with reconstructed direction less than $10^{\mathrm{o}}$ of the Earth limb are rejected,
thus reducing contamination from Earth's gamma-ray albedo.
W Com was observed about 3 degrees off-axis with respect to the boresight and a
3.7$\sigma$ excess (pre-trials) of events $>$100 MeV is found from 12 (03:00 UT) to 13 (03:00)
June 2008, corresponding to a flux of $(90\pm32) \times 10^{-8}$ ph~s$^{-1}$~cm$^{-2}$.
It should be mentioned that 
this flux is roughly a factor of 1.5 higher than the highest flux detected by the
Energetic Gamma Ray Experiment Telescope (EGRET; \cite{Hartman1999}) onboard the
Compton Gamma Ray Observatory and significantly higher than the weekly 
averaged peak flux of $(17.2\pm3.5)\times10^{-8}$ ph s$^{-1}$ cm$^{-2}$ reported by 
the Large Area Telescope onboard the Fermi Gamma-Ray Space Telescope during its first three months of 
operation \citep{Abdo-2009a}.
No excess $>3\sigma$ is
found in the rest of the observing period and upper limits are obtained; 
results can be found in Table \ref{table:AGILEobservations} and Figure \ref{fig-lightcurve}.

SuperAGILE, the hard X-ray imager onboard AGILE (18-60 keV, \cite{Feroci2007})
observed the source for a net exposure time of 253 ks.
The source position in the orthogonal SuperAGILE reference system is $\sim$ (3,0) deg,
which means that the exposed area is close to the full on-axis effective area \citep{Feroci2007}.
W Com is not been detected with SuperAGILE, and
we estimate a 3$\sigma$ upper limit in the 20-60 keV energy of 6 mCrab $\simeq 6.9 \times 10^{-11}$ 
erg~cm$^{-2}$~s$^{-1}$ (assuming a photon index of $\Gamma=2.1$).

\subsection{\emph{Swift} and \emph{XMM-Newton}: X-ray observations}

Observations of W Com with the \emph{Swift} satellite 
\citep{Gehrels-2004} were taken on 2008 June 7-9. 
All \emph{Swift} X-ray telescope (XRT) data \citep{Burrows-2005} are reduced 
using the HEAsoft 6.5 package. Event files are calibrated 
and cleaned following the standard filtering criteria 
using the \emph{xrtpipeline} task and applying the most 
recent Swift XRT calibration files. All data were taken in 
Photon Counting (PC) mode, with grades 0-12 selected over 
the energy range 0.3-10 keV. Due to photon pile-up in the core 
of the point spread function (PSF) at rates larger than 0.5 counts 
s$^{-1}$ (PC mode), the source events are extracted from an 
annular region with an inner radius of 3 pixels and an outer 
radius of 30 pixels (47.2 arcsec). Background counts are 
extracted from a 40 pixel radius circle in a source-free region. 
Ancillary response files are generated using the \emph{xrtmkarf} 
task, with corrections applied for the PSF losses and CCD defects. 
The response matrix Version 11 from the XRT calibration files is 
applied. To ensure valid $\chi^{2}$ minimization statistics during
spectral fitting, the extracted XRT energy spectra are re-binned to 
contain a minimum of 20 counts in each bin.
The spectra can be described by a single power-law convolved with galactic 
and local absorption.
Table
\ref{table:Swiftobservations} summarizes the observations along with the best fit
model parameters.

W Com was observed by the XMM-Newton Observatory \citep{Jansen-2001} between
2008 June 14 and June 18 over three consecutive orbits. The three
observations comprise data from the EPIC detector (0.2-10~keV) in Small Window
mode.
The data have been analysed using SASv7.1
\citep{Gabriel-2004}. Several filtering criteria have been applied to the EPIC
data, including filtering for time periods of high background activity
following the standard procedure, and filtering only for single- and double-pattern
events for EPIC-pn and single to quadruple for EPIC-MOS, as well as
including only %events with quality FLAG=0. 
events with good quality (quality FLAG=0).
For the spectral analysis, circular
source and annular background extraction regions centered on the source are
selected by maximizing the signal-to-noise ratio. The spectra are re-binned in
order not to oversample the intrinsic energy resolution of the EPIC cameras by
a factor not more than 3, while making sure that each spectral channel
contains at least 25 background-subtracted counts. 
This allows the use of the
$\chi^2$ quality-of-fit estimator to find the best fit model. Fits are
performed in the 0.2-10~keV energy range simultaneously for the three EPIC
cameras, where the systematic difference between the EPIC cameras is below
$\sim$5$\%$ in normalization. For the spectral analysis and fitting procedure
XSPEC v12.4 \citep{Arnaud-1996} is used. The data can be best described 
similar to the XRT data by a
single power-law convolved with galactic and local absorption. Table
\ref{table:XMMobservations} summarizes the observations along with the best fit
model parameters.

\begin{figure}
\plotone{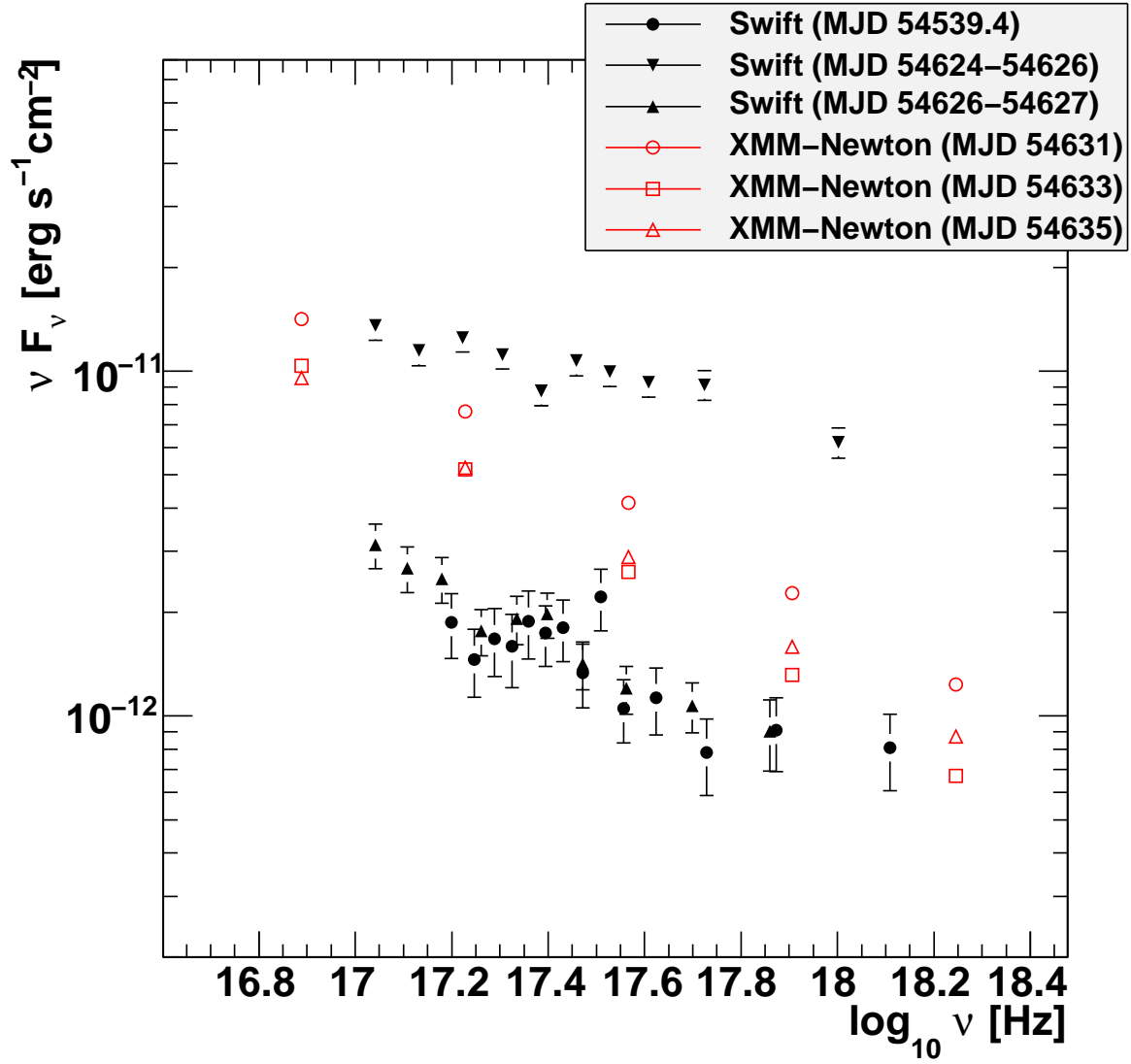}
\caption{\label{fig-X-raySED}
Spectral energy distribution for five different X-ray measurements with \emph{Swift XRT} and \emph{XMM-Newton EPIC} for
2008 June 7-18 and the Swift XRT measurements in March 2008 (MJD 54539.4).
}
\end{figure}

The measurements reveal strong variability in X-rays on time scales of much less than one day.
Fig \ref{fig-lightcurve} (panel C)  and \ref{fig-X-raySED} show that the X-ray flux changed by
a factor of two during the VHE high state on MJD 54625.
This is comparable to observations of W Com with \emph{Beppo}Sax in 1998 by \cite{Tagliaferri-2000},
where flux variations of a factor of three in less than 5 hours is reported.
The X-ray flux during the VHE low state of June 2008 is very similar to the X-ray activity measured during the 
detection of W Com in March  2008 (see Figure \ref{fig-X-raySED}).

\subsection{Optical, Near-IR, UV and Radio Observations}

Eight optical, one near-IR and three radio observatories contributed data sets to this campaign;
see Table \ref{table:Observatories} for an overview.
The majority of the observatories are part of the GLAST-AGILE Support Program (GASP, see \cite{Villata-2008}),
 a subgroup of the Whole Earth Blazar Telescope 
(WEBT\footnote{http://www/oato.inaf.it/blazars/webt}).
In the period considered here, optical observations of W Com were carried out at the following observatories: 
Abastumani, Crimean, Roque de los Muchachos (KVA), Talmassons, Torino (for details concerning
these observatories, see references provided by WEBT), San Pedro Martir, Northern Optical Telescope
(NOT\footnote{http://www.not.iac.es/}) and Sapienza University (Italy).
Magnitude calibration is obtained with respect to the photometric sequence by \cite{Fiorucci-1996}.
Near-infrared ($JHK$) data were acquired at the AZT-24 telescope at Campo Imperatore Observatory (Italy).
\emph{Swift} UV/Optical Telescope (UVOT) \citep{Roming-2005} observations were taken in the photometric bands of 
\emph{UVW1} (centered at 2600 \AA), \emph{UVM2} (centered at 2246 \AA), and \emph{UVW2} (centered at 1928 \AA) \citep{Poole-2008}. 
The \emph{uvotsource} tool is used to extract counts
from the UVOT, correct for 
coincidence losses, apply background subtraction, and calculate 
the source flux. The standard 5 arcsec radius source aperture is 
used, with a 20 arcsec background region. 

At radio frequencies, data at 43 GHz were taken  with the 32 m antenna at Noto \citep{Bach-2007}, at 
14.5 GHz with the 26 m telescope of the UMRAO \citep{Aller-2003}, and at 36.8 GHz with the 13.7 m
Mets\"ahovi radio telescope \citep{Terasranta-1998}.

Data reduction of the optical and radio data
followed standard methods and procedures, and we refer to the above papers for details.
The near infrared, optical and UV data are corrected for absorption
 in our Galaxy using the dust maps of \cite{Schlegel-1998} and the extinction curve of \cite{Cardelli-1989}.  
Since the blazar is observed in a bright state (see Section \ref{sec:Model}), a host
galaxy contribution has not been subtracted.

\section{Modeling and Discussions}
\label{sec:Model}

\begin{figure}
\plotone{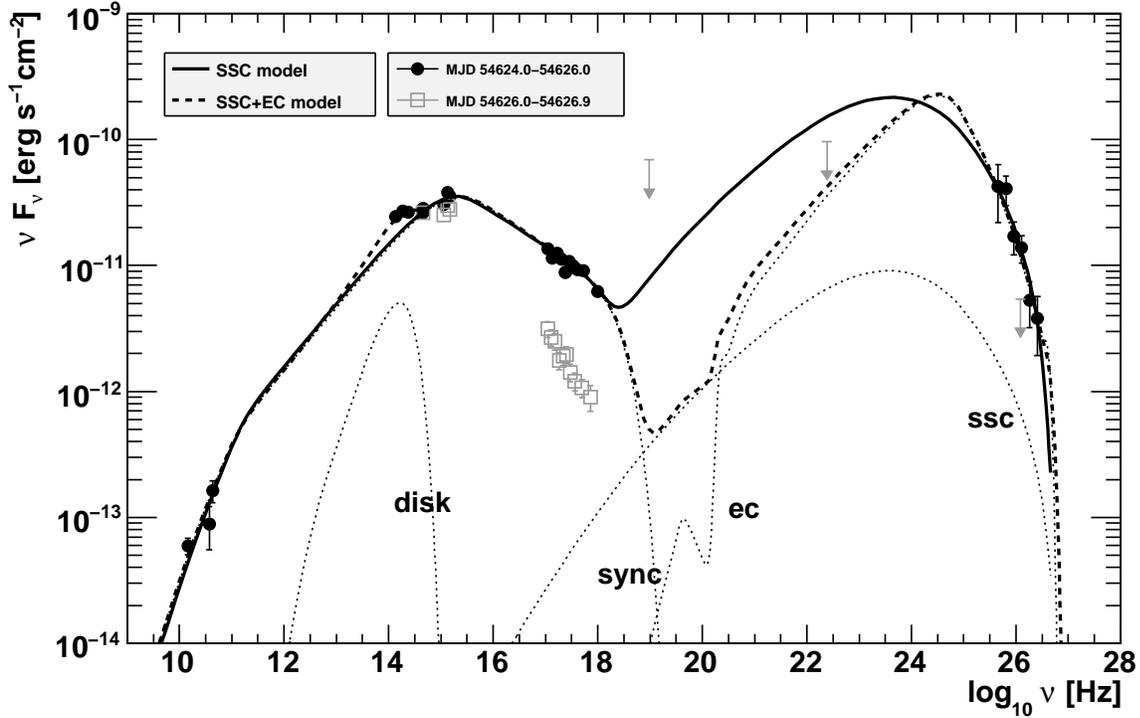}
\caption{\label{fig-SEDMJD54624}
High-state spectral energy distribution of W Com for MJD 54624 to 54626 including VERITAS, Swift XRT/UVOT,
optical and radio data (filled circular markers). The averages of the optical, NIR and radio fluxes
calculated over the time range
from MJD 54610 to 54645 are shown here. 
Downward-pointing arrows indicate upper flux limits  (99\% confidence level) \citep{Helene-1983}.
For comparison, 
the VERITAS, AGILE and Swift XRT/UVOT data for MJD 54626 to 54626.9
are shown as grey open squares and grey downward pointing arrows.
Results from  synchrotron-self-Compton (SSC) and external-Compton (SSC+EC) models are shown as
continuous and dashed lines, respectively (see text for details). The different components (disk, sync=synchrotron, ec, ssc) of the SSC+EC models are indicated by
dotted lines.
}
\end{figure}

The single-epoch SEDs for two different time intervals are shown in 
 Fig. \ref{fig-SEDMJD54624}.
The broad-band SEDs of W Com show  double-humped structures, as found
in all known gamma-ray blazars.
The photon-production mechanism in these objects are successfully modeled by leptonic 
(e.g. \cite{Boettcher-2002}, \cite{Ghisellini-1996}) 
and hadronic model (e.g. \cite{Boettcher-2002a}, \cite{Muecke-2003}, \cite{Aharonian-2000}).
The data presented here have been modeled using a leptonic one-zone jet model.
For this purpose, a quasi-equilibrium
version of the model described in \cite{Boettcher-2002} is
adopted. 
In this model, the observed electromagnetic radiation
is interpreted as originating from ultrarelativistic electrons
(and positrons) in a spherical emission region of co-moving
radius $R_B$, which is moving with a relativistic speed
$\beta_{\Gamma} c$, corresponding to the bulk Lorentz factor
$\Gamma$. 
Lacking more detailed constraints on the viewing
angle $\theta$ between the jet direction and the line of sight, 
we fix $\theta$ to be the superluminal angle, for which the
bulk Lorentz factor $\Gamma$ equals the Doppler factor $D =
\left( \Gamma [1 - \beta_{\Gamma} \cos\theta] \right)^{-1}$,
which determines the Doppler shift of photon energies and
relativistic boosting of intensities. We note that our results
mainly depend on $D$ so that alternative combinations of
$\Gamma$ and $\theta$ yielding the same Doppler factor as the
ones used in our model calculations are also possible, although
minor differences in the flux of the external-Compton
emission with respect to other radiation components (see
below) would result (see, e.g., Dermer 1995). 

In our calculations, the size of the emission region is
constrained by the shortest observed variability time
scale $\delta t_{\rm var, min}$ through $R_B \le c \delta 
t_{\rm var, min} \, D / (1 + z)$. In the optical and X-rays \citep{Boettcher-2002a},
variability down to time scales of a few hours has been
observed, limiting the blob radius to $R_B \le 10^{15}
(\delta t_{\rm var, min}/{\rm hr}) \, (D/10)$~cm.

Ultrarelativistic electrons are assumed to be instantaneously
accelerated into a power-law distribution in electron energy,
$E_e = \gamma m_e c^2$, as $Q(\gamma) = Q_0 \gamma^{-q}$ with
a low- and high-energy cutoff $\gamma_1$ and $\gamma_2$, respectively.
An equilibrium between this particle injection, radiative cooling,
and escape of particles from the emission region yields a temporary
quasi-equilibrium state described by a broken power-law. The
particle escape is parameterized through an escape time
scale parameter $\eta > 1$ as $t_{\rm esc} = \eta R/c$.  The
balance between escape and radiative cooling will lead to a
break in the equilibrium particle distribution at a break
Lorentz factor $\gamma_b$, where $t_{\rm esc} = t_{\rm cool} 
(\gamma)$. The cooling time scale is evaluated self-consistently
taking into account synchrotron, synchrotron-self-Compton (SSC) 
and external Compton (EC) cooling. Depending on whether $\gamma_b$ 
is greater than or less than $\gamma_1$, the system will be in
the slow cooling or fast cooling regime. In the fast cooling
regime ($\gamma_b < \gamma_1$), the equilibrium distribution
will be a broken power-law with $n(\gamma) \propto \gamma^{-2}$
for $\gamma_b < \gamma < \gamma_1$ and $n(\gamma) \propto
\gamma^{-(q+1)}$ for $\gamma_1 < \gamma < \gamma_2$. In the
slow cooling regime ($\gamma_b > \gamma_1$), the equilibrium
distribution will be $n(\gamma) \propto \gamma^{-q}$ for
$\gamma_1 < \gamma < \gamma_b$ and $n(\gamma) \propto
\gamma^{-(q+1)}$ for $\gamma_b < \gamma < \gamma_2$.
The number density of injected particles is normalized
to the resulting power in ultrarelativistic electrons
propagating along the jet, 

\begin{equation} 
L_e = \pi R_e^2 \, \Gamma^2 \beta_{\Gamma} \, c \, m_e c^2 
\, \int\limits_1^{\infty} \gamma n(\gamma) d\gamma.
\label{Le}
\end{equation}

The magnetic field $B$ in the emission region is 
a free parameter. 
The corresponding Poynting flux along the
jet is $L_B = \pi R_e^2 \, \Gamma^2 \beta_{\Gamma} \, c \, u_B$,
with the magnetic energy density $u_B = B^2/(8\pi)$.
For each model calculation, the resulting equipartition parameter,
$e_B = L_B/L_e$, is evaluated. Modeling results of a large number
of blazars, in particular flat-spectrum radio quasars,
have shown that leptonic models can achieve
reasonable fits with the emission region being close to equipartition,
typically $0.1 \lesssim e_B \lesssim 1$. However, there is no a 
priori argument which would dictate quasi-equipartition. Therefore,
while we disfavor possible fit results with $e_B$ far from unity,
we can not strictly rule out such scenarios.

Once the quasi-equilibrium particle distribution in the emission
region is calculated, our code evaluates the radiative output from 
synchrotron emission, SSC, and EC emission self-consistently with 
the radiative cooling rates. For the EC component, we assume an
external radiation field which is isotropic in the stationary
AGN rest frame and can be approximated by a thermal blackbody with
peak frequency $\nu_{\rm ext}$ and radiation energy density $u_{\rm ext}$.
The latter two quantities are free model parameters. The
direct emission from this external radiation field is added to
the emission from the jet to yield the total model SED which we 
fit to the observations.

In all model calculations the luminosity distance to W Com
 has been calculated using
standard $\Lambda$-CDM cosmology with $\Omega_m=0.3$ and 
$\Omega_{\Lambda}=0.7$.
Absorption of high-energy gamma rays by the extragalactic background light 
is taken into account using the model of \cite{Franceschini-2008}.

We fit the VERITAS flare detection and high X-ray state (MJD 54624.0 -- 54626.0)
 with
a pure SSC model, i.e., without any external radiation fields,
and with a model with an EC component. 
A Doppler factor of 20 (i.e., $D = \Gamma = 20$) consistent with
all observational constraints, and well in the range of Doppler
factors commonly adopted in other blazar modeling works, allowed
acceptable fits to the SEDs. We therefore fixed $D = \Gamma =
20$ for the remainder of the fitting procedure. 

For a pure SSC fit, the free parameters were thus (1) $L_e$, the 
injection power of ultrarelativistic electrons into the emission 
region, (2,3) $\gamma_1$ and $\gamma_2$, the cutoffs of the injected 
electron distribution, (4) $q$, the injection spectral index, (5) $B$, 
the magnetic field, (6), $R_B$, the radius of the emission region, 
and (7) $\eta$, the particle escape time scale parameter. The injection 
spectral index is tightly constrained by the observed X-ray energy 
spectral index $\alpha = q/2$, since electrons emitting synchrotron 
radiation in the X-ray regime are always above the critical Lorentz 
factor $\gamma_b$. The radius of the emission region is constrained
through the minimum variability time scale of a few hours, as
mentioned above. Together with the value of the magnetic field,
the low-energy cutoff $\gamma_1$, determines the location of the
synchrotron and gamma-ray peaks in the SED, while the high-energy
cutoff $\gamma_2$ influences the location of the high-energy cutoffs
of the SED, in particular the synchrotron component. The cutoff of
the SSC component is, in addition, strongly influenced by Klein-Nishina
effects. Parameters of the SSC fit shown in Fig.  \ref{fig-SEDMJD54624}
 are listed in Table \ref{FITparameters}.
 
 No SSC model fit was possible with the emission region being close to equipartition.
Since there is virtually no observational
constraint on the high-energy emission in the low (MJD 54626) and intermediate
X-ray state (MJD 54631), we could choose a low injection power and
relatively high magnetic field to achieve a synchrotron peak flux comparable to the flaring
state, but at a much lower SSC flux. Such a choice of parameters allowed us to bring the
system close to equipartition. However, almost any positive detection either in the Fermi
or the VHE gamma-ray range could rule out this interpretation. In the SSC interpretation,
the most significant difference between the various states consists of a change in the electron
injection spectral index q from 2.55 in the flaring state to 3.50 and 3.40 in the low and
intermediate state.

For a model with an EC component, two more parameters need to be specified: (8) the 
peak frequency $\nu_{\rm ext}$ and the energy density $u_{\rm ext}$ of the external 
radiation field. 
As with the SSC model, the electron spectral index $q$ is tightly
constrained through the X-ray spectral index, while the variability time scale
constrains the radius of the emission region. In order to avoid the problem of
required large injection powers (to obtain a high SSC flux) and accordingly
small magnetic fields (not to overpredict the synchrotron flux), the VHE gamma-ray
emission can be interpreted as external-Compton emission. In order for Comptonization
of an external radiation field to be efficient out to gamma-ray energies of
$E \gtrsim E_{\rm VHE} = 300$~GeV, the external radiation field has to peak at 
energies $E_{\rm ext} \lesssim (m_e c^2)^2/E_{\rm VHE} \sim 0.9$~eV, i.e., in the
near-infrared. Therefore, line emission from a putative broad-line region (for
which there is no evidence in W~Comae), would have a too high
photon energy  characteristic to serve efficiently as a source photon field
for EC scattering to produce an IC spectrum with peak
energy near the VHE gamma-ray band.
It is therefore more likely that infrared emission,
e.g., from a near-nuclear dust torus, dominates the external radiation field
responsible for EC emission at VHE gamma rays. We find that an external radiation
field peaking at $\nu_{\rm ext} = 1.5 \times 10^{14}$~Hz can, at the same time,
serves as an efficient source for EC emission and explains the slight
near-IR bump in the SED of W~Comae.
This bump could also be due to the host galaxy, and future observations
of variability of the IR component or very high-resolution imaging 
are required to break this degeneracy. 
 With the assumption of such an external
radiation field, acceptable fits to each of the states of W~Comae can be achieved
within a factor of $\sim 3$ of equipartition. The parameters of our SSC+EC fit
are listed in Table \ref{FITparameters}.

\section{Conclusions}

W Com belongs to the IBL class of blazars, a group with a now-growing number of VHE-detected blazars.  
Other blazars detected with VHE gamma rays that are not of the HBL class
include the IBLs 3C 66A  \citep{Acciari2009}  and PKS 1424+240 \citep{Ong-2009},
low-frequency-peaked BL
Lac objects (LBLs) such as BL Lacertae \citep{Albert-2007},
and the flat-spectrum radio quasar 3C279 \citep{Albert-2008}.
 In this article, we described a second VHE flare measured from W Com by VERITAS. 
 The object was detected by VERITAS at a significance level of 10.3 standard deviations during  2008 June 7-8. 
 The VERITAS observations triggered a multiwavelength campaign including AGILE gamma-ray, 
 \emph{Swift} and \emph{XMM-Newton} X-ray, UV, optical and radio observations. 
 We have carried out extensive modeling of the SED of W Com constructed from this contemporaneous
  multiwavelength data set, 
  using a leptonic model considering
  synchrotron, SSC, and external-Compton emission. 
The SED can be modeled by a
simple leptonic SSC model, but the wide separation of
the peaks in the SED requires a rather low ratio of the magnetic
field to electron energy density of $\epsilon_B = 2.3\times 10^{-3}$.
 The SSC+EC model returns magnetic field parameters closer
to equipartition, providing a satisfactory description of
the broadband SED. These findings are similar to the results
obtained from the first W Com VHE flare reported by \cite{Acciari2008b}.

The strong variability of W Com at X-ray and gamma-ray energies
on time scales of days or less shows that only truly
contemporaneous data can provide serious constraints on
the various emission models. 
Future observations with VERITAS and the \emph{Fermi Gamma-Ray Space Telescope} 
should provide even more detailed data to better resolve
the short variability timescales, 
helping to further constrain model calculations.

\acknowledgments

This research is supported by grants from the US Department
of Energy, the US National Science Foundation, and the Smithsonian
Institution, by NSERC in Canada, by Science Foundation
Ireland, and by STFC in the UK.
We acknowledge the
excellent work of the technical support staff at the FLWO and
the collaborating institutions in the construction and operation
of the instrument.
Financial support by the Italian Space Agency through contract ASI-INAF I/088/06/0 is acknowledged.
Support of UMRAO from NSF and University of Michigan is acknowledged.
NVE and VML acknowledge support from RFBR grant 09-02-00092.
The GASP-WEBT consortium is acknowledged.
We acknowledge the efforts of the Swift team
for providing the UVOT and XRT observations.
This work was partially supported by NASA through XMM-
Newton Guest Observer Program award No. NNX08AD67G 
and the Swift Guest Investigator Program award No. NNX08AU13G.
The Mets\"ahovi team acknowledges the support from the Academy of Finland. 

{\it Facilities:} \facility{VERITAS}, \facility{AGILE}, \facility{Swift}, \facility{XMM-Newton}, \facility{GASP-WEBT}

\clearpage

\begin{deluxetable}{cccc}
\centering
 \tablecolumns{2}
\tablewidth{0pt}
\tablecaption{Observatories contributing to the presented data set.
\label{table:Observatories}}
\tablehead{
\colhead{Waveband} &
\colhead{Observatory}  &
\colhead{Frequency/Band/} &
\colhead{MJD Range} \\
\colhead{} &
\colhead{} &
\colhead{Energy Range} &
\colhead{} 
}
\startdata
Radio & UMRAO & 14.5 GHz  & 54630-54633\\
& Mets\"ahovi & 36.8 GHz &  54623-54634 \\
& Noto & 43 GHz & 54611 \\
NIR/Optical/UV & NOT & U/B/V/R/I & 54636 \\
& Tuorla & R & 54622 - 54645 \\
&  Abastumani & R & 54617 - 54637 \\
& Sapienza University & R & 54627 - 54634 \\
& San Pedro Martir & R & 54620 - 54624 \\ 
& KVA & R & 54626 - 54633 \\
& Crimean & R & 54623 - 54627 \\
& Talmassons & R & 54628 \\
& Torino & R & 54630 \\
& Campo Imperatore & J/H/K & 54627 - 54633 \\
& Swift UVOT & U/B/V/UV  & 54625 - 54626 \\
X-ray & Swift XRT & 0.3-10 keV & 54625 - 54626 \\
& SuperAGILE & 20-60 keV & 54626 - 54630\\
& XMM-Newton EPIC & 0.2-10 keV & 54631 - 54635 \\
HE Gamma-ray & AGILE GRID & 30 MeV - 30 GeV  & 54626 - 54632  \\
VHE Gamma-ray & VERITAS & 0.1-30 TeV & 54624 - 54626 \\
\enddata
\end{deluxetable}

\begin{deluxetable}{cccccc}
\centering
\tablecolumns{6}
\tablewidth{0pt}
\tablecaption{
Details of VERITAS observations of W Com on 2008 June 7-9.
The energy threshold for fluxes and upper flux limits (99\% confidence level; assuming a photon index of $\Gamma=3.68$)
is 200 GeV. Errors are given at the 1$\sigma$ level.
\label{table:VERITASobservations}}
\tablehead{
\colhead{MJD} &
\colhead{elevation} &
\colhead{observation}  &
\colhead{average pedestal} & 
\colhead{significance} &
\colhead{flux or }\\
\colhead{} &
\colhead{range} &
\colhead{time}  &
\colhead{variations\tablenotemark{a}} &
\colhead{(pre-trials)} &
\colhead{upper flux limit} \\
\colhead{} &
\colhead{} &
\colhead{[min]} &
\colhead{[dc]} &
\colhead{[$\sigma$]} &
\colhead{[cm$^{-2}$ s$^{-1}$]} 
}
\startdata
54624.16 - 54624.23 & 53-73$^{\mathrm{o}}$ & 100.2 & 7.8-8.0 & 8.9 & $(5.0\pm0.8)\times 10^{-11}$   \\
54625.17 - 54625.24 & 49-68$^{\mathrm{o}}$ & 100.2 & 8.1-9.7 & 7.9 & $(6.2\pm1.2)\times 10^{-11}$   \\
54626.18 - 54626.20 & 59-60$^{\mathrm{o}}$ & 32.0  & 12.2-12.3\tablenotemark{b}  & -1.0 & $<3.21\times 10^{-11}$ \\
\enddata

\tablenotetext{a}{The average pedestal variation in digital counts [dc] indicates the background light level.
Values of 6.5 to 6.8 are typical for regular observations of extragalactic targets on moonless nights.
All observations presented here are taken
in moonlight conditions.}
\tablenotetext{b}{Data taken with increased pixel (PMT) trigger threshold (at 70 mV CFD trigger threshold instead of the regular 50 mV).}
\end{deluxetable}

 \begin{deluxetable}{ccc}
 \centering
 \tablecolumns{3}
\tablewidth{0pt}
\tablecaption{Details and results of the AGILE GRID observations of \mbox{W Com} on 2008 June 9-15. 
The energy threshold for fluxes and upper flux limits (99\% confidence level; assuming a photon index of $\Gamma=2.1$)
is 100 MeV.
Errors are given at the 1$\sigma$ level.
 \label{table:AGILEobservations}}
 \tablehead{
\colhead{MJD} &
\colhead{significance} &
\colhead{flux or} \\
\colhead{} &
\colhead{(pre-trials)} &
\colhead{upper flux limits } \\
\colhead{} &
\colhead{} &
\colhead{ [cm$^{-2}$ s$^{-1}$] }
}
\startdata
54626.75 - 54629.12 & $<3\sigma$ & $<60\times10^{-8}$  \\
54629.12 - 54630.12 & $3.7\sigma$ & $(90\pm34)\times10^{-8}$ \\
54630.12 - 54632.50 & $<3\sigma$ & $<55.5\times10^{-8}$ \\
\enddata
\end{deluxetable}

 \begin{deluxetable}{cccc}
 \centering
 \tablecolumns{4}
\tablewidth{0pt}
\tablecaption{Details and results of the \emph{Swift/XRT} observations of \mbox{W
 Com} 2008 June 7-9. The galactic N$_{H,Gal}$ has been
  fixed to a value of 1.88$\cdot$10$^{20}$ cm$^{-2}$.
 The redshift of the source was assumed to be 0.102. 
Errors are given at the 1$\sigma$ level.
 \label{table:Swiftobservations}}
 \tablehead{
\colhead{MJD}          &
\colhead{Exposure}     &
\colhead{Photon index}     &
\colhead{Flux F$_{2-10 \mathrm{keV}}$} \\
\colhead{} &
\colhead{[ksec]} &
\colhead{$\Gamma$} &
\colhead{[10$^{-12}$ ergs cm$^{-2}$ s$^{-1}$]}
}
\startdata
54624.97 - 54624.98 & 0.52 &  $2.49\pm 0.19$ &  $3.90\pm 0.97$ \\
54625.04 - 54625.05 & 0.84 &  $2.71\pm 0.15$  & $3.70\pm 0.76$ \\
54625.11 - 54625.12 & 1.38 &  $2.55\pm 0.09$  & $4.75\pm 0.55$ \\
54625.17 - 54625.20 & 2.51 &  $2.36\pm 0.05$  & $9.33\pm 0.74$ \\
54625.24 - 54625.27 & 2.47 & $2.59\pm 0.07$ & $4.62 \pm0.37$ \\
54626.11 - 54626.21 & 5.07 & $2.69\pm 0.10$  & $1.00 \pm 0.18$ \\
\enddata
\end{deluxetable}
 
 \begin{deluxetable}{ccccc}
 \centering
 \tablecolumns{5}
\tablewidth{0pt}
\tablecaption{Details and results of the XMM-Newton observations of \mbox{W
 Com} 2008 June 14-18. The galactic N$_{H,Gal}$ has been
  fixed to a value of 1.88$\cdot$10$^{20}$ cm$^{-2}$ as obtained from
  \cite{Dickney-1990}. The redshift of the source was assumed to be
  0.102. Errors are given at the 1 sigma level.
 \label{table:XMMobservations}}
 \tablehead{
\colhead{MJD}          &
\colhead{Exposure}     &
\colhead{N$_{H}$}      & 
\colhead{Photon index}     &
\colhead{Flux F$_{2-10 \mathrm{keV}}$}\\
\colhead{ } &
\colhead{[ksec]} &
\colhead{[10$^{20}$ cm$^{-2}$]} &
\colhead{ $\Gamma$} &
\colhead{[10$^{-12}$ ergs cm$^{-2}$ s$^{-1}$]}
}
\startdata
54631.50 - 54631.55 & 28.0 & 2.20$^{+0.09}_{-0.09}$ & 2.79$^{+0.01}_{-0.01}$ & 2.69$^{+0.02}_{-0.02}$ \\
54633.15 - 54633.17 & 16.0 & 1.39$^{+0.14}_{-0.13}$ & 2.88$^{+0.02}_{-0.02}$ & 1.53$^{+0.03}_{-0.02}$ \\
54635.14 - 54635.16 & 11.0 & 1.05$^{+0.16}_{-0.15}$ & 2.77$^{+0.02}_{-0.02}$ & 1.89$^{+0.03}_{-0.03}$ \\
\enddata
\end{deluxetable}

\begin{table}[hbtp]
\caption[]{Parameters of SSC and SSC+EC fits to the SEDs of W Com on MJD 54624.0 --
54626.0.}
\label{FITparameters}
\begin{center}
\begin{tabular}{cccc}
\hline
{Parameter} & {Symbol} & {SSC} & {SSC+EC} \\
\hline
Doppler factor 		& $D$	     & $20$ & 20  \\
Electron power  [erg~s$^{-1}$]	& $L_e$	     & $3.4\times 10^{44}$ & $5.7\times 10^{43}$  \\
Blob radius 	  [cm]     & $R_b$	     & $3 \times 10^{15}$ & $10^{16}$   \\
Low-energy cutoff	& $\gamma_1$ & $9 \times 10^3$ & $8 \times 10^3$   \\
High-energy cutoff      & $\gamma_2$ & $2.5 \times 10^5$ & $3 \times 10^5$  \\
Electron injection index & $q$	     & 2.55 & 2.55 \\
Magnetic field   [G]       & $B$        & 0.24  & 0.35  \\
B-field equipartition parameter      & $e_B$	& $2.3 \times 10^{-3}$ & 0.32 \\
Electron escape time scale parameter & $\eta$	& 300 & 300   \\
Minimum variability time scale  [hr] & $\delta t_{\rm var, min}$ & 1.5  & 5.1 \\
External radiation peak frequency [Hz] & $\nu_{\rm ext}$ & -- & $1.5 \times 10^{14}$ \\
External radiation energy density [erg~cm$^{-3}$] & $u_{\rm ext}$ & -- & $2.4 \times 10^{-4}$ \\
\hline
\end{tabular}
\end{center}
\end{table}

%%%%%%%%%%%%%%%%%%%%%%%%%%%%%%%%%%%%%%%%%%%%%%%%%%%%%%
%%%%%%%%%%%%%%%%%%%%%%%%%%%%%%%%%%%%%%%%%%%%%%%%%%%%%%
%%%%%%%%%%%%%%%%%%%%%%%%%%%%%%%%%%%%%%%%%%%%%%%%%%%%%%
%% FIGURES
%%%%%%%%%%%%%%%%%%%%%%%%%%%%%%%%%%%%%%%%%%%%%%%%%%%%%%
%%%%%%%%%%%%%%%%%%%%%%%%%%%%%%%%%%%%%%%%%%%%%%%%%%%%%%
%%%%%%%%%%%%%%%%%%%%%%%%%%%%%%%%%%%%%%%%%%%%%%%%%%%%%%

\end{document}